\documentclass[aps,prl,twocolumn,showpacs,floatfix,a4paper]{revtex4}
\usepackage{graphicx}
\usepackage[intlimits]{amsmath}

\begin{document}

\title{Entanglement properties and quantum phases
for a fermionic disordered one dimensional wire with attractive interactions}
\author{Richard Berkovits}
\affiliation{Department of Physics, Jack and Pearl Resnick Institute, Bar-Ilan
University, Ramat-Gan 52900, Israel}

\begin{abstract}
A fermionic disordered one dimensional wire in 
the presence of attractive interactions
is known to have two distinct phases: A localized and a superconducting one
depending on the strength of interaction and disorder. The localized region may
also exhibit a metallic behavior if the system size is shorter than the 
localization length. Here we show that the superconducting phase has
a distinct distribution of the entanglement entropy and
entanglement spectrum distinct from the metallic regime.
The entanglement entropy distribution is strongly asymmetric with 
L\'evy alpha stable distribution (compared to the Gaussian metallic 
distribution), and the entanglement level spacing distribution is unitary
(compared to orthogonal). Thus, entanglement properties may reveal properties
which can not be detected by other methods.
\end{abstract}


\maketitle

%\section{Introduction}
%
%The statistical properties of mesoscopic metallic systems have been 
%at the center of attention for the last three decades \cite{akkermans07}.
%Phenomena such as universal conductance fluctuations (UCF)\cite{lee85},
%angular correlations in coherent multiple scattering \cite{berkovits94},
%and level number fluctuations in a metallic grain \cite{altshuler86}
%have garnered much interest.

In the last decade the application of concepts from quantum information, such
as entanglement entropy (EE)\cite{amico08},
took center stage in understanding physical phenomena in condensed matter
physics.
One of the reasons for this interests
is that EE is deeply connected to quantum phase
transitions (QPT). The EE quantifies
the entanglement in a many-body system by dividing it into two
regions A and B. 
For a system in a pure state $|\Psi\rangle$, the
entanglement between regions A and B is measured by the EE $S_{A/B}$ 
connected to the eigenvalues of the reduced density matrix of area A, $\rho_A$ 
or B, $\rho_B$.
It is expected that non-local properties, such as the EE, may provide
a different perspective beyond the traditional point-point correlations
and local order parameters \cite{song12}.

Specifically, $\rho_A$ is defined as:
$\rho_{A}={\rm Tr}_{B}|\Psi\rangle\langle \Psi |$, where the
degrees of freedom of region B are traced out.
The eigenvalues of the matrix $\lambda_i^{A}$ are used to calculate the 
EE:
\begin{eqnarray}
S_{A}=- \sum_i \lambda_i^{A} \ln \lambda_i^{A} .
\label{ee}
\end{eqnarray}
Recently it has been understood that one can 
utilize the information enclosed in the full spectrum
of $\{\lambda^A_i\}$.
These eigenvalues are used to construct the entanglement spectrum (ES)
$\{\varepsilon_i^A=- \ln \lambda_i^{A}\}$ \cite{haldane08}.
For one-dimensional (1D) systems, the area of the boundary between regions
A and B is fixed and thus the EE should not depend on the size of region A.
Nevertheless a logarithmic dependence of the form
\cite{holzhey94,vidal03,calabrese04,korepin04}:
\begin{eqnarray} \label{ee_av}
S(L_A,L) = \frac{1}{6}
\ln\left(\sin\left(\frac{\pi L_A}{L}\right)
\right)+c,
\end{eqnarray}
where $L_A$ is the length of region A and $L$ is the samples length,
is expected in the metallic (clean) regime.
Several years ago Li and Haldane \cite{haldane08} came
up with an intriguing conjecture regarding the 
connection between the ES 
and the excitation spectrum of a many-body state.
They suggested that the low-energy ES 
shows a precise correspondence to the true many-particle
spectrum of region A. 
The properties of the EE and 
the ES may be used to identify 
phase transitions 
in disordered many-body properties
\cite{prodan10,gilbert12,berkovits12,chen12,mondragon13,chu13,berkovits14,pouranvari14,romer15}.

In this letter we will use the EE and ES in order to
investigate the nature of different phases of 
fermionic disordered 1D systems with attractive interactions.
Electron-electron interactions in 
1D systems are parametrized by the Luttinger parameter $K$
\cite{apel_82,giamarchi88}.
For non-interacting systems $K=1$, while for attractive 
interactions $K>1$. 
When both disorder and interaction are present,
an extended metallic (with superconducting correlations) phase
is expected once attractive interactions are strong enough, i.e.
$K>1.5$ \cite{giamarchi88,kane97,sohn97,mirlin07}.
This stems from the 
renormalization group scaling 
of the localization length 
\begin{eqnarray} \label{xi_u}
\xi = (\xi_0)^{1/(3-2K)},
\end{eqnarray} 
where $\xi_0$ is the non interacting localization length.
Thus, for $K=1.5$ the localization length 
diverges, and one transits from the localized
to the extended regime.
Indeed, it has been numerically demonstrated that with strong
enough attractive interactions in the usual Anderson model
\cite{schmitt98,schuster02,carter05,chu13}
metal-like behavior emerges, although no evidence of superconducting
correlation have been numerically demonstrated.

For disordered system
one must consider the EE  behavior over an ensemble of different
realizations of disorder. It has been demonstrated
in Ref. \cite{berkovits12} 
that the median EE for
$L_A<\xi$ follows the metallic logarithmic behavior 
(Eq. (\ref{ee_av})), while for $L_A>\xi$ it saturates. 
In principal this could be used to decide in what phase (localized
or metallic) the system is in \cite{chu13}. Nevertheless, in a realistic
numerical study this strategy is fraught with problems, since the 
localization length grows fast as function of $K$ (Eq. (\ref{xi_u})),
and easily outgrows any  finite system length
$L$. Once $\xi \gg L$ a finite
system will show metallic behavior although it is in the localized
regime. For brevity, we shall refer to the 
$K>1.5$ regime as superconducting, and to the finite sample $K<1.5$ 
regime as localized or metallic according to whether $L>\xi$ or $\xi>L$.

In this Letter we will show that the full distribution of the
EE shows a distinct behavior between the metallic and superconducting regime, 
although the median EE is essentially identical
in both regimes. 
The EE distribution changes from a Gaussian in the metallic regime 
to a very asymmetric L\'evy alpha stable distribution with ``fat tails''
in the superconducting regime.
The ES level spacing (ESLS) distribution 
fits the Gaussian orthogonal distribution
(GOE) expected from interacting many particle system
\cite{berkovits94,berkovits96,pascaud98,berkovits99,song00,oganesyan07,prodan10,gilbert12}
in the metallic regime, while it changes to a 
Gaussian unitary distribution (GUE)
associated with superconducting excitations \cite{berkovits95,altland97}
in the superconducting regime.
Thus, the EE and ESLS distributions are able to characterize the phase of 
the system, where other methods fail.

In this letter
we consider a spinless 1D electrons wire of size $L$ 
with {\it attractive} nearest neighbor
interactions and on-site disordered potential.
The system's Hamiltonian is given by:
\begin{eqnarray} \label{hamiltonian}
H &=& 
\displaystyle \sum_{j=1}^{L} \epsilon_j {\hat c}^{\dagger}_{j}{\hat c}_{j}
-t \displaystyle \sum_{j=1}^{L-1}({\hat c}^{\dagger}_{j}{\hat c}_{j+1} + h.c.) \\ \nonumber
&+& U \displaystyle \sum_{j=1}^{L-1}({\hat c}^{\dagger}_{j}{\hat c}_{j} - \frac{1}{2})
({\hat c}^{\dagger}_{j+1}{\hat c}_{j+1} - \frac{1}{2}),
\end{eqnarray}
where $\epsilon_j$ is the on-site energy, which 
is drawn from a uniform 
distribution $[-W/2,W/2]$,
${\hat c}_j^{\dagger}$ is the creation 
operator of a spinless electron at site $j$,
and $t=1$ is the
hopping matrix element.
The interaction strength is $U<0$, 
and a background is included.
For the non-interacting Anderson model the 
system is localized with a localization
length $\xi_0 \approx 105/W^2$
\cite{romer97}. 
Here the Luttinger parameter $K(U)=\pi/[2 \cos ^{-1} (-U/2)]$ 
\cite{g_formula,giamarchi03}. 
For non-interacting electrons $K(U=0)=1$.
For attractive interactions $K>1$ and $\xi$ increases as $U$ becomes more
negative. For $U=-1$, $K=1.5$ and the localization length according
to Eq. (\ref{xi_u}) diverges. 
Thus, below $U<-1$ the system is expected to be delocalized. 
At $U=-2$ it goes through another phase transition
to a phase separated state and is insulating again.
Indeed numerically \cite{schmitt98,schuster02,carter05,chu13}, 
this system is known to exhibit extended
behavior for a range of attractive interaction strength
centered around $U=-1.5$ and not too strong disorder $W<1.5$ 

The density matrix renormalization group (DMRG) \cite{white92,dmrg}.
is a very accurate numerical method for calculating
the ground state of disordered interacting 1D system and
for the calculation of the reduced density matrix.
We calculate the EE for three length $L=300,700,1100$
and different values of $L_A=10,20,\ldots,L-10$, for $400,200,100$
realizations of disorder for the corresponding system length.
Specifically, we calculate the normalized EE of 
the j-th realization at a given $L_A$,
$s_j(L_A) = S_j(L_A)/\langle S(L_A) \rangle$, 
where $\langle S(L_A) \rangle$ is
the average EE over the different realizations. 
Since the distribution of the
EE is very similar for different values of $L_A$ as long as $L_A$
is not to close to the edge we accumulate the distribution of
the normalized EE, $P(s)$, in the range of $L/4<L_A<3L/4$.

\begin{figure}
\includegraphics[width=8cm,height=!]{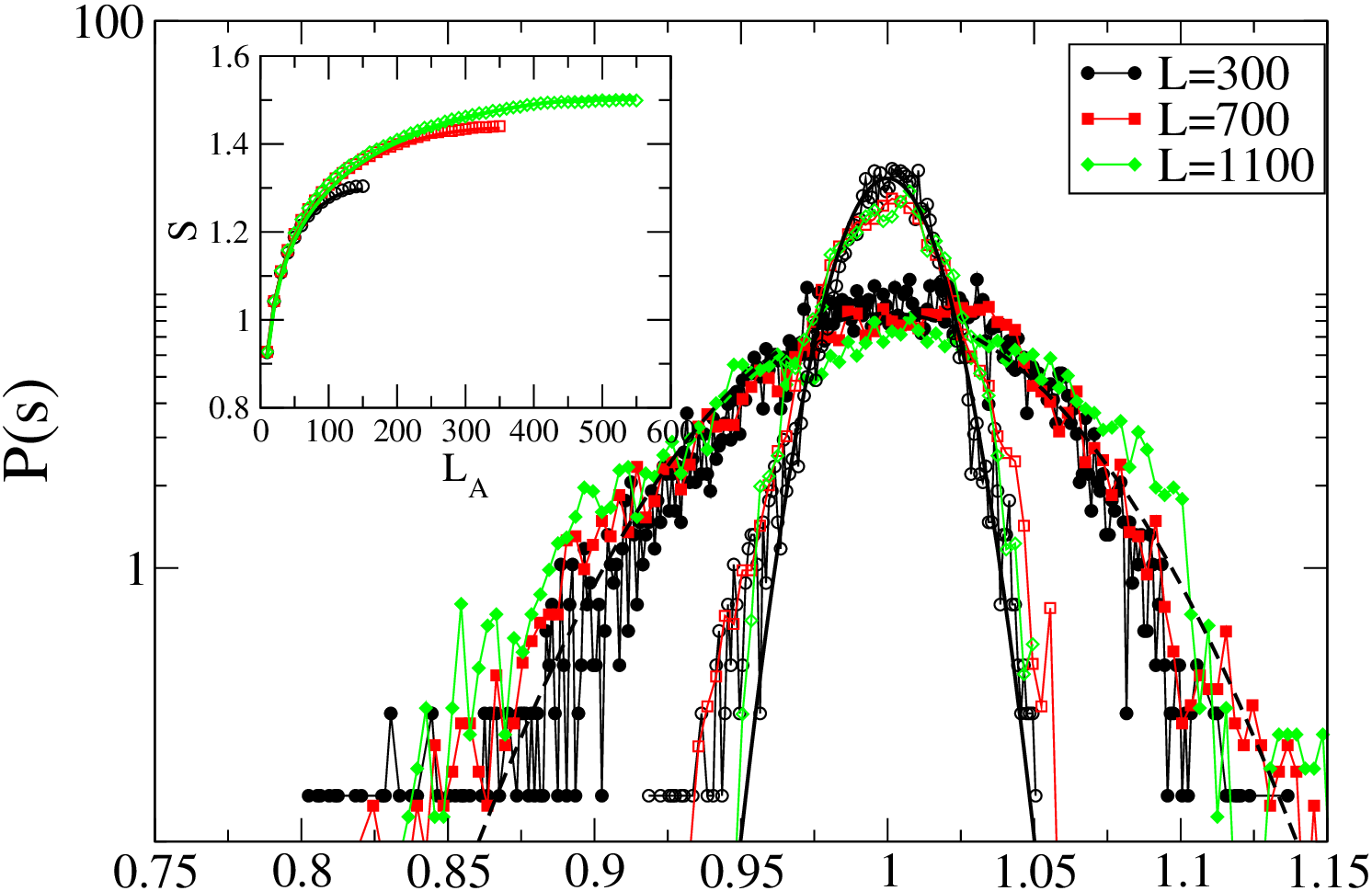}
\caption{\label{fig0}
(Color online)
The distribution $P(s)$ of the normalized EE 
(see text) for different length $L=300$ (black, circles), $L=700$
(red, squares), $L=1100$ (green, diamonds), and disorder
strength $W=0.3$ (empty symbols), $W=0.7$ (full symbols). 
A fit to a Gaussian with a width which depends on
the disorder, $\sigma(W)$,
is depicted by the continuous lines.
(inset) The median EE as function of $L_A$. The symbols
correspond to the numerical results, the curves to Eq. (\ref{ee_av}).
}
\end{figure}

Let us begin by discussing the EE distribution for
$U=-0.7$ ($K(U=-0.7)= 1.3$) for which the system
is in the metallic regime, i.e., localization length much larger than
sample length. In Fig. \ref{fig0} 
we present the distribution for two values of disorder ($W=0.3,W=0.7$),
corresponding to the non-interacting 
localization length $\xi_0(W=0.3) \sim 1200$ 
and $\xi_0(W=0.7) \sim 200$, thus $\xi(W=0.3,U=-0.7) \sim 5 \cdot 10^{7}$, 
and $\xi(W=0.7,U=-0.7) \sim 6 \cdot 10^{5}$.
First, we examine the median
value of the EE as function of $L_A$ (here we could use the
average which is almost equal to the median, but for the sake of uniformity
with the upcoming results we use here the median) and compare it
with the expression of the EE for a clean system described in
Eq. (\ref{ee_av}).
As can be seen in the inset of Fig. \ref{fig0}
it fits quite well. Thus, the median EE follows closely
the expected behavior of a clean (metallic) system.
The distribution $P(s)$ for all three length and two disorder strength
is plotted in Fig. \ref{fig0}. As has been discussed in
Ref. \cite{berkovits12a}, for $L \ll \xi$, we expect the distribution 
to be Gaussian (i.e. $(\sqrt{2 \pi} \sigma)^{-1}\exp(-(x-1)^2/2 \sigma^2$)
centered at the average. 
This is indeed seen in Fig. \ref{fig0} (up to a slight skewness of the tails), 
as well as the
fact that the width, $\sigma$ of the Gaussian is almost independent of
$L$. On the other hand it is clear that the disorder strength $W$ 
does affect $\sigma$. As might be expected, The weaker is 
the disorder, the narrower is the width of the Gaussian.

\begin{figure}
\includegraphics[width=8cm,height=!]{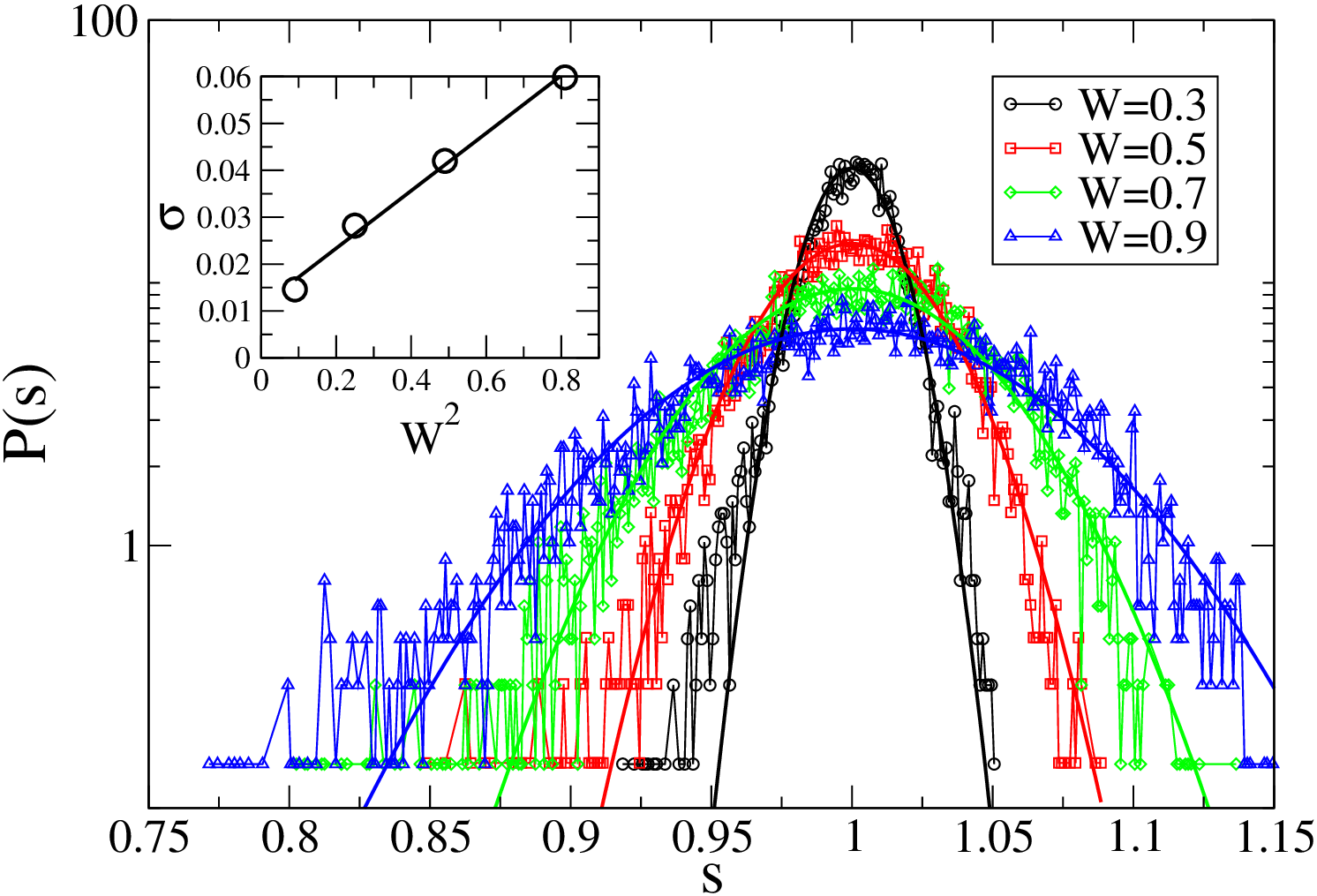}
\caption{\label{fig0a}
(Color online)
The distribution $P(s)$ of the normalized EE 
for a fixed wire length ($L=300$)
and different disorder strength: $W=0.3$ (black circles), 
$W=0.5$ (red squares), $W=0.7$ (green diamonds), $W=0.9$ (blue triangles).
A fit to a Gaussian of width, $\sigma(W)$ (depicted
in the inset) is indicated by the continuous lines.
(inset) The Gaussian width $\sigma$ as function of the disordered
strength square $W^2$.
}
\end{figure}

The dependence of the Gaussian width on $W$ is examined
in Fig. \ref{fig0a}. We keep the length and interaction strength
fixed ($L=300$, $U=-0.7$) while varying $W$.
Even for the strongest disorder $\xi(W=0.9,U=-0.7) \sim 2 \cdot 10^{5}$, 
is much larger than the sample
length. $P(s)$ remains Gaussian for all disorder strength.
Moreover, as can be seen from the inset $\sigma(W) \propto W^2$. 

\begin{figure}
\includegraphics[width=8cm,height=!]{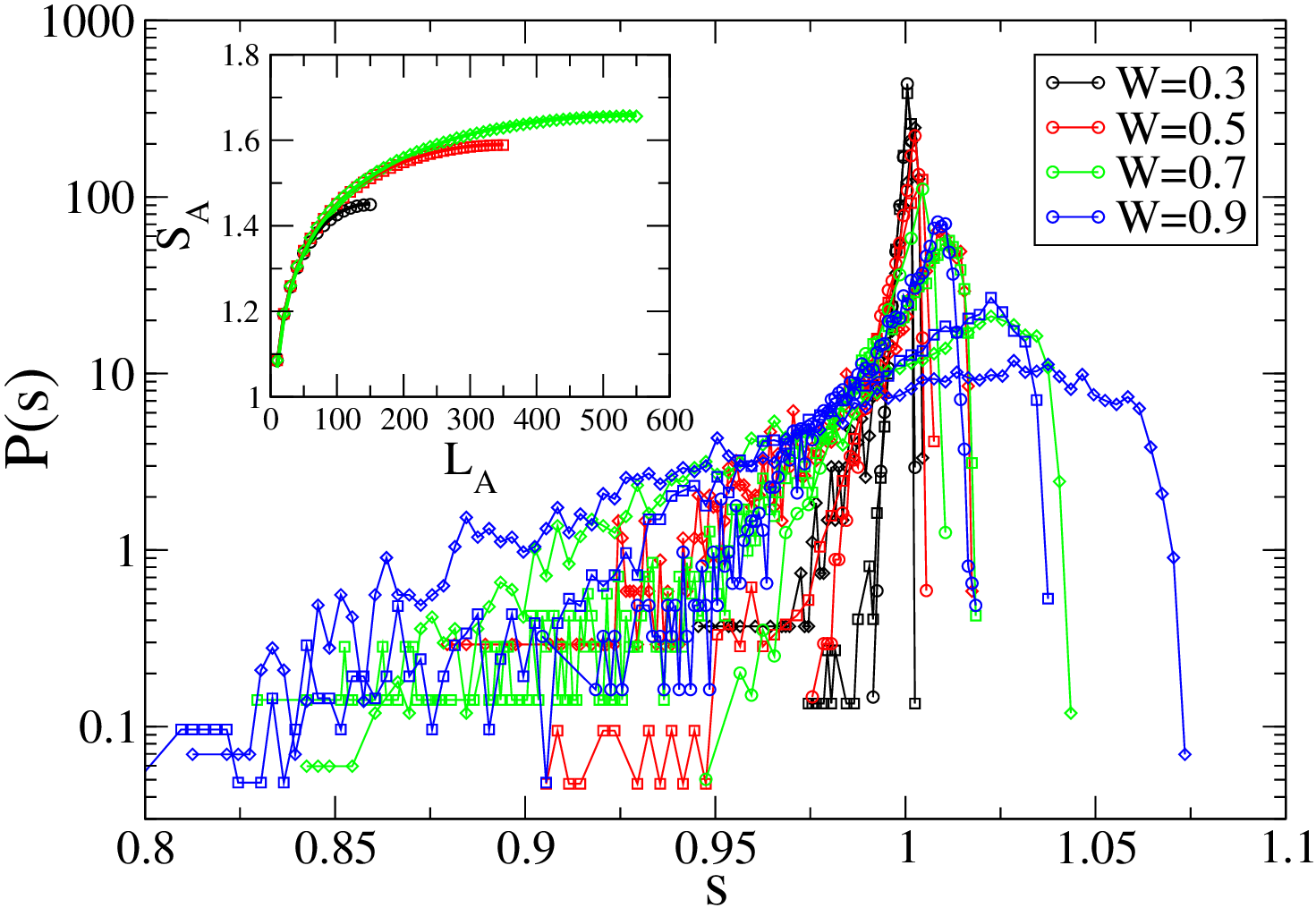}
\caption{\label{fig1} (Color online)
The distribution $P(s)$ of the normalized EE deep in the
superconducting regime ($U=-1.5$) for different  
strength of disorder ($W=0.3$, black symbols,
$W=0.5$, red symbols, $W=0.7$, green symbols,
$W=0.9$, blue symbols), and 
sample length $L=300$ (circles), $L=700$
(squares), $L=1100$ (diamonds). 
(inset) The median EE as function of $L_A$. The symbols
correspond to the numerical results, the curves to Eq. (\ref{ee_av}).
}
\end{figure}

What happens to the EE distribution in the superconducting regime? Specifically,
we concentrate on the extended regime with $U=-1.5$
($K(U=-1.5)= 2.3$), different
values of disorder $W=0.3,0.5,0.7,0.9$ 
corresponding to a non-interacting mean free path
$\xi_0 \sim 1200,400,200,130$ and different sample length
$L=300,700,1100$. This regime of the
parameter space is deep in the superconducting regime.
We present the distribution $P(s)$ for all length and interaction
strength  in Fig. \ref{fig1}.
It is obvious that $P(s)$ is completely different than
in the metallic regime (Figs. \ref{fig0},\ref{fig0a}).
In all cases the distribution
is strongly asymmetric. The typical value of the EE is close to
its maximum value, and the probability of measuring an EE
larger than the typical value is rather small and falls
off abruptly. On the other hand, there is a high probability 
for measuring values of the EE below the typical value.
For smaller values of the EE the distribution has a very long
tail. Another difference is the strong dependence on the sample length $L$
even at the same value of disorder $W$. 

First, one concludes that similarly to the case of the EE
of a localized system ($L > \xi$)
\cite{berkovits12,berkovits12a,berkovits14},
the average EE is not a correct description of the
superconducting regime typical EE. The EE is more 
appropriately represented
by the median value. In the inset of Fig. \ref{fig1}
the median
value of the EE as function of $L_A$ is compare
to Eq. (\ref{ee_av}). As can be seen it fits
quite well.

\begin{figure}
\includegraphics[width=8cm,height=!]{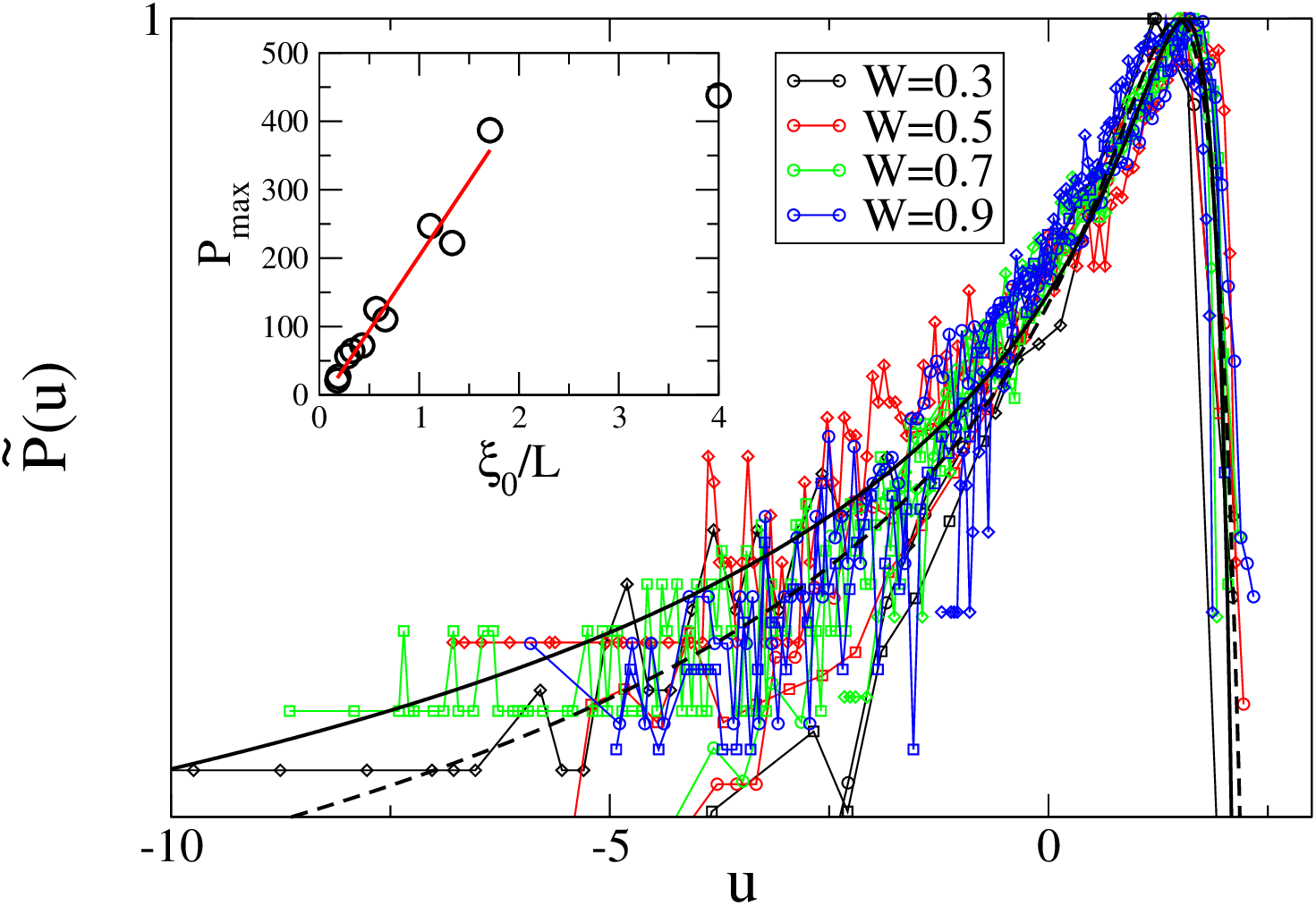}
\caption{\label{fig1a} (Color online)
The data presented in Fig. \ref{fig1}
scaled by the parameter $u$ (see text). 
The continuous (dashed) black curve 
is the L\'evy stable distribution with
$\alpha=1,\beta=-1,\gamma=0.285,\delta=0.8$,
($\alpha=1.2,\beta=-1,\gamma=0.28,\delta=0.52$).
(inset) The maximum of the distribution $P(s)$ ($P_{\rm max}$) depicted
in Fig. \ref{fig1}
as function of $\xi_0/L$. A linear behavior in the regime $\xi_0/L<2$ is
indicated by the line.
}
\end{figure}

As can be seen in Fig. \ref{fig1a},
the distribution is universal and may be scaled by the function
$\tilde P(u)$, here
$u=1+(s-1)*P_{\rm max}$, and the distribution is normalized
$\tilde P(u) = P(u)/P_{\rm max}$,  
where $P_{\rm max}$ is the value of $P(s)$ at the maximum. 
In the inset of Fig. \ref{fig1a} , the values of $P_{\rm max}$
vs. $\xi_0/L$ is depicted. Up to the ballistic regime, 
(i.e.,for $\xi_0 \gg L$), a linear relation seems to hold.
The distribution $\tilde P(u)$ may be described rather well
by a L\'evy alpha stable distribution, which 
is a natural extension for the central limit theorem to the case that the
identically-distributed random variables have no finite variance, and is
used to describe distributions with ``fat tails''.
The L\'evy distribution 
\begin{eqnarray} \label{levy}
f(x,\alpha,\beta,\gamma,\delta)=\frac{1}{\pi} {\rm Re}
\int_0^\infty e^{it(x-\mu)}e^{-(\gamma t)^\alpha(1-i\beta\Phi)}dt,
\end{eqnarray} 
(where $\Phi=\tan(\pi \alpha/2)$, except for $\alpha=1$ where
$\Phi=-(2/\pi)\log(t)$),
is defined
by four parameters. The stability index $0 \leq \alpha <2$ characterizes the
asymptotic behavior of the tails $|x|^{-1-\alpha}$ (except for $\alpha=2$), 
the skewness parameter
$-1\geq \beta \leq 1$, $\gamma$ is a scale factor and
$\delta$ controls the location of the maximum \cite{burnecki12}.
In general $f$ is not analytical, except for special cases,
such as $f(x,\alpha=2,\beta,\gamma=\sigma/\sqrt{2},\delta=\bar{x})$,
equal to a Gaussian of width $\sigma$.
We plot $f(u,\alpha=1,\beta=-1,\gamma=0.285,\delta=0.8)$, and 
$f(u,\alpha=1.2,\beta=-1,\gamma=0.28,\delta=0.52)$ in Fig. \ref{fig1a}.
The skewness is clearly maximal ($\beta=-1$), while fitting $\alpha$ depends
very much on the tail region, which has the largest numerical uncertainty.
Nevertheless, the main part of the distribution is
evidently fitted well by $1<\alpha<1.2$.

Thus, although the median of the EE (or its average) does not
give us a clear signature whether the system is in the metallic regime
or in the
superconducting one, the distribution can differentiate between the
regimes. Further work is needed in order to follow how the distribution
changes as one goes through the phase transition and is the transition
point characterized by a special distribution. A hint may be found in
Ref. \cite{chu13}, where the variance of the EE is plotted through the
transition. It seems that the variance becomes smaller in the metallic side
as one approaches the transition and then grows again in
the superconducting regime. This is in line with a crossover from
a Gaussian distribution to a fat tail one. 

%The shorter samples ($L=300,700$) are in the
%ballistic regime, $L<\ell$, and $P(s)$ in both cases is essentially
%identical. For the longer smaples ($L=1100$), a pronaunced change in the
%distribution $P(s)$ is noticable. 
%, although for $L=1100$ ($L \sim \ell$)
%a longer tail is observed.

\begin{figure}
\includegraphics[width=8cm,height=!]{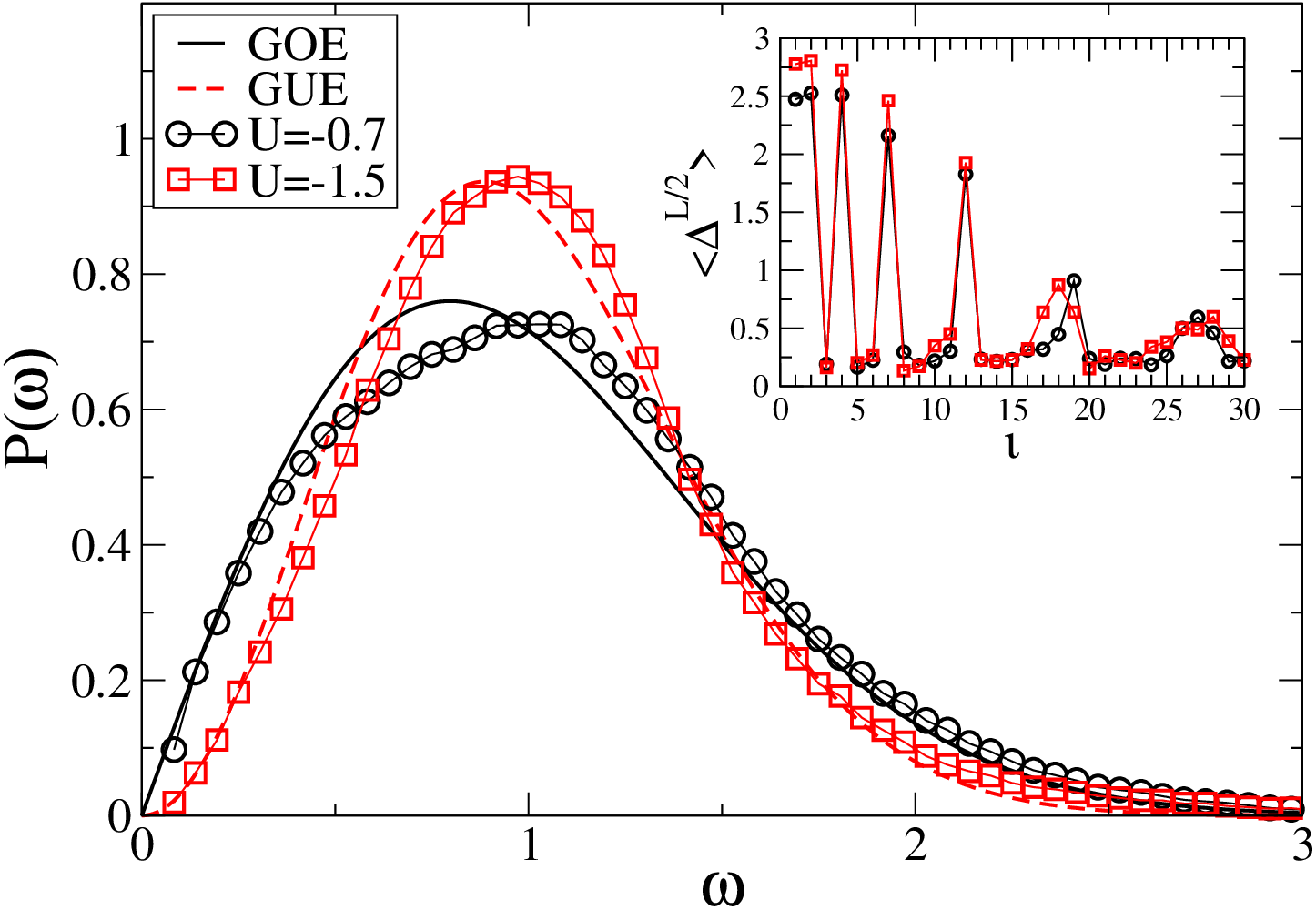}

\caption{\label{fig3}
(Color online)
The ESLS distribution $P(\omega)$ as function of the
normalized level spacing $\omega$. The numerical results for 
the metallic ($U=-0.7$, black circles) and superconducting 
($U=-1.5$, red squares) regimes for $L=700$ are indicated.
The black (red dashed) line correspond to the GOE (GUE) distribution.
Inset: The average level spacing $\langle \Delta^{N_A}_i \rangle$ as function
of level number for both interactions.
}
\end{figure}

In calculating the ESLS one must take into account that
different sectors of the reduced density matrix
with a fixed number of electrons do not couple with each other.
Therefore, for each eigenstate of $\rho_A$ one should calculate
both $\lambda_i$ and $N^A_i$ (the number of particles in the region $A$).
Calculating $N^A_i$ does not add to the complexity of 
the DMRG code \cite{song12}.
Thus, each eigenvalue of the reduced density matrix has two indexes
 $\lambda^{N_A}_i$, which are translated into the 
entanglement Hamiltonian eigenvalues  \cite{haldane08}
$\varepsilon^{N_A}_i=-\ln(\lambda^{N_A}_i)$.
It makes sense to calculate the ESLS 
only between eigenvalues belonging to the same
number sector. Thus, 
$\Delta^{N_A}_i= \varepsilon^{N_A}_{i+1} - \varepsilon^{N_A}_i$,
and $omega^{N_A}_i=\Delta^{N_A}_i/\langle \Delta^{N_A}_i \rangle$.
The average spacing for $U=-0.7$ and
$U=-1.5$ for $N_A=L/2=350$ as function of $i$
is depicted in the inset of Fig. \ref{fig3}.
$\langle \Delta^{L/2}_i \rangle$ only slightly depends
on $U$. There are several anomalously large values appearing for the first
few spacings.
The distribution $P(\omega)$ is accumulated over all low-lying spacings 
($\varepsilon^{N_A}_i<20$), which are not
anomalously large (i.e., $\langle \Delta^{N_A}_i \rangle<1$)
for all values of $N_A$, and $L/4<L_A<3L/4$.
The distributions for the metallic ($U=-0.7$) and superconducting
($U=-1.5$) regimes are presented in Fig. \ref{fig3}.
It can be seen that for the metallic case
the numerical results fit quite well the GOE distribution
($P_{\rm GOE}(s)=(\pi s/2)\exp(-\pi s^2/4)$)
expected for the excitations of interacting many particle systems
\cite{berkovits96,pascaud98,berkovits99,song00,oganesyan07,prodan10,gilbert12}.
On the other hand, in the superconducting regime the distribution
follows the GUE distribution 
($P_{\rm GUE}(s)=(32 s^2/\pi^2)\exp(-4 s^2/\pi)$). 
Indeed, superconducting correlations 
lead to exotic statistics of the excitations close to
the Fermi energy \cite{altland97}, which are quite similar to GUE
for higher energies.
Thus, also
the ESLS distribution shows a significant difference
between the metallic and superconducting phases, although for the
length considered here ($L=700 \ll \xi=6 \cdot 10^{5}$) both regions 
are extended. 

To conclude, the EE and ESLS distributions show a distinct
behavior, depending if the system is in a metallic (system length much
smaller than the localization length) or a superconducting (localization
length diverges) regime, 
for a disordered one dimensional spinless electron system
with attractive interactions. While the metallic regime shows a rather
expected behavior, e.g., Gaussian distribution of the EE and GOE ESLS 
distribution, the superconducting EE shows an asymmetric
distribution and GUE ESLS. Thus, the entanglement properties
encode details of the underlying phase of the system which may elude
other measures, which may be useful for detecting new phases for
different systems. The emergence of a ``fat tail'' distribution for the
EE in the superconducting regime is fascinating and deserves further
study.

%>>>>>>>>>>>>>>>>>>>>>>>>>>>>>>>>>>>>>>>>>>>>>>>>>>>>>>>>>>

\begin{acknowledgments}
I would like to thank B. Altshuler, I. Aliener and V. Oganesyan,
for useful discussions.
Financial support from the Israel Science Foundation (Grant 686/10) is
gratefully acknowledged.
\end{acknowledgments}

\end{document}